\documentclass{iaus}
\usepackage{graphicx}

\title[Eternally Collapsing Objects] 
{Masses of Radiation Pressure Supported Stars in Extreme Relativistic Realm}

\author[Abhas Mitra]   
{Abhas Mitra }%

\affiliation{Theoretical Astrophysics Section,
BARC, Mumbai-400085, India \break email: amitra@barc.gov.in\\[\affilskip]}

\pubyear{2006}
\volume{238}  
\pagerange{001--999}
\date{??? and in revised form ???}
\jname{Black Holes: from Stars to Galaxies -- across the Range of Masses}
\editors{V. Karas \& G. Matt, eds.}
\begin{document}

\maketitle

\begin{abstract}
We discuss that in the extreme relativistic limit, i.e., when $z >>1$, where $z$ is the surface gravitational redshift,
there could be radiation pressure supported and dominated stars with arbitrary gravitational mass, high or low. Such Objects
are called Eternally Collapsing Objects (ECOs).  ECOs
are practically as compact as Schwarzschil Black Holes (BH) and, observationally, are likely to be mistaken as BHs. Further
since any object undergoing continued collapse must first become an ECO before becoming a true BH state charcterized by $M=0$,
the observed BH Candidates are ECOs.
\end{abstract}

\section{Introduction}
Let the ratio of radiation pressure to gas pressure in a star be $x= p_r/p_g$, and let the ratio of energy densities
of radiation and rest mass be $y= \rho_r/\rho_0$. Then it follows that, within a factor of $2.0$, $y \sim \alpha ~z$,
where $\alpha = L/L_{ed}$ (\cite{1}). Here $L$ is the luminosity of the star and $L_{ed}$ is the maximal or Eddington luminosity
of the same. It is also seen that $\alpha$ increase with $z$ and saturates to a maximal value of $\alpha \approx 1$.
The conventional known ``stars'' have $z <<1$, i.e., they are Newtonian objects. In such a case  
$x \approx 0.2 (M/ M_\odot)^{1/2} ~ \mu^{-1}$
where $\mu$ is the chemical composition and $M_\odot$ is solar mass (\cite{2}). In order to have a ``radiation pressure supported star'', one must have $x \gg 1$.
Then the above equation demands that $M \gg M_\odot$. Such stars are known as (Newtonian) Supermassive Stars. But this picture changes
dramatically when we allow for the likely occurrence of extreme relativistic situations with $z \gg 1$ when (\cite{2}) 
\begin{equation}
x \approx 0.25 ~z~ \left({M\over M_\odot}\right)^{1/2} ~ \mu^{-1}
\end{equation}
Because of the presence of $z$ on the RHS, now a radiation supported quasistatic state ($x \gg 1$) can occur for arbitrary value of $M$. For instance
an object with $M =10^{10} M_\odot$ with a value of $z =100$ will have $x \gg 1$ and on the other hand, a fireball of energy 1 TeV created by two colliding
elementary particles in a lab collider with $M =10^{-43} M_\odot$  could attain a radiation dominated quasistatic state with $z> 10^{25}$! In both the
cases, the radius of the object would be practically same as those of  corresponding (supposed) BHs. Such a large value of $z$ may sound unphysical. But, actually, it is not
so because continued collapse is supposed to proceed to a true BH state with $z=\infty$ and all finite numbers including $10^{25}$ are {\em infinitely
smaller than} $\infty$. It has been also shown that, as  physical gravitational collapse must be radiative (\cite{3}) and $z$ keeps on increasing relentlessly during continued collapse, the quanta of collapse generated
radiation quanta must eventually move in {\em almost} closed circular orbits and be quasitrapped even before an {\em absolute} trapping
would occur by the formation of trapped surfaces or Event Horizon (EH). It is this quasitrapping of radiation which causes $x \gg 1$ and also $y \gg 1$.
Since such states correspond to $\alpha \approx 1$, ECOs are formed much before a true BH state ($z=\infty$) would be formed. Since the life time
of this relativistic radiation dominated state is $t= 5. 10^{8} ~(1+z)$ yr and $t \to \infty$ as $z \to \infty$ ECOs are {\em eternally collapsing} (\cite{4}).
\section{Discussion}
As the ECOs evolve, their EOS $\rho \to 3 p$ and the gravitational mass $M = \int (\rho -3p) d\Omega \to 0$   where $d \Omega$ is  proper
volume  element (\cite{5}). Consequently, the eventual BH mass $M=0$ and thus the observed BHs   with $M >0$, either discussed in this conferences or elsewhere, cannot be true BHs and must be ECOs. This conclusion  that true BHs have $M \equiv 0$ has been also obtained by using the basic differential geometry property that proper spacetime 4-volume is coordinate independent (\cite{6}; \cite{7}) and non-occurrence of trapped surfaces and EH (\cite{8}, \cite {9}). Since  astrophysical
plasama is endowed with intrinsic magnetic field the super compact ECOs must have strong intrinsic magnetic fields and hence they are also called
Magnetospheric ECOs or MECOs. In contrast a true uncharged BH with an EH has no intrinsic magnetic field. Thus MECOs can be observationally distinguished from true BHs by using this physical property. And there are strong circumstantial evidences that the BH Candidates in (i) X-ray binaries
(\cite{10}, \cite{11}, \cite{12}) or the (ii) the compact object at the galactic centre (\cite{13}) are MECOs with physical surfaces rather than any EH. Most importantly,
detailed microlensing aided mapping of the central engine of the one of the most well studied quasar has directly revealed that it is a MECO and
not a BH (\cite{14}).
The observed magnetic field of MECOs will however be smaller by a factor of ($1+z$) and could be well below  pulsar values. Because of joint
effect of gravitational redshift and time dilatation, the observed luminosity of the ECOs would be also also extremely small, $L = 1.3 \times 10^{38} (M/M_\odot) ~(1+z)^{-1}$ erg/s even though they are shinging at thir  respective Eddington values. This feeble quiesent ECO quiesent radiation could be in the microwave  range and is yet to be detected for any individual case.
Since ECO formation is a generic effect due to inevitable quasi-trapping of collapse generated radiation, continued collapse cannot give rise to any
mysterious phase transition resulting in negative pressure or ``gravstars''. Since Chandrasekhar mass limit pertains to {\em cold} objects, it is irrelevant for
{\em hot} ECOs whose local mean temperature is $T = 600 (M/M_\odot)^{-1/2}$ MeV.

\end{document}